\documentstyle[12pt,aaspp4]{article}

\begin{document}

\title{A STUDY  OF THE  SHADOWING OF GALACTIC COSMIC RAYS BY THE SUN 
IN A QUIET PHASE OF SOLAR ACTIVITY WITH  THE TIBET AIR SHOWER ARRAY}

\author{ M.~Amenomori\altaffilmark{1}, S.~Ayabe\altaffilmark{2}, 
        Caidong\altaffilmark{3},       Danzengluobu\altaffilmark{3},
        L.K.~Ding\altaffilmark{4},     Z.Y.~Feng\altaffilmark{5}, 
        Y.~Fu\altaffilmark{6},          H.W.~Guo\altaffilmark{3}, 
        M.~He\altaffilmark{6},          K.~Hibino\altaffilmark{7},
        N.~Hotta\altaffilmark{8},       Q.~Huang\altaffilmark{5},
        A.X.~Huo\altaffilmark{4},       K.~Izu\altaffilmark{9},
        H.Y.~Jia\altaffilmark{5},       F.~Kajino\altaffilmark{10}, 
        K.~Kasahara\altaffilmark{11},   Y.~Katayose\altaffilmark{12},
        Labaciren\altaffilmark{3},      J.Y.~Li\altaffilmark{6},
        H.~Lu\altaffilmark{4},          S.L.~Lu\altaffilmark{4},
        G.X.~Luo\altaffilmark{4},       X.R.~Meng\altaffilmark{3}, 
        K.~Mizutani\altaffilmark{2},     J.~Mu\altaffilmark{13}, 
        H.~Nanjo\altaffilmark{1},       M.~Nishizawa\altaffilmark{14},
        M.~Ohnishi\altaffilmark{9},     I.~Ohta\altaffilmark{8},
        T.~Ouchi\altaffilmark{7},       Z.R.~Peng\altaffilmark{4},       
        J.R.~Ren\altaffilmark{4},
        T.~Saito\altaffilmark{15},      M.~Sakata\altaffilmark{10},
        T.~Sasaki\altaffilmark{10},     Z.Z.~Shi\altaffilmark{4},
        M.~Shibata\altaffilmark{12},    A.~Shiomi\altaffilmark{9}, 
        T.~Shirai\altaffilmark{7},      Y.~Suga\altaffilmark{10},
        H.~Sugimoto\altaffilmark{16},
        K.~Taira\altaffilmark{16},      Y.H.~Tan\altaffilmark{4},
        N.~Tateyama\altaffilmark{7},    S.~Torii\altaffilmark{7}, 
        T.~Utsugi\altaffilmark{2},      C.R.~Wang\altaffilmark{6},
        H.~Wang\altaffilmark{4},        
        X.W.~Xu\altaffilmark{4,9},        Y.~Yamamoto\altaffilmark{10},
        G.C.~Yu\altaffilmark{3},        A.F.~Yuan\altaffilmark{3}, 
        T.~Yuda\altaffilmark{9, 17},        C.S.~Zhang\altaffilmark{4},
        H.M.~Zhang\altaffilmark{4},     J.L.~Zhang\altaffilmark{4},
        N.J.~Zhang\altaffilmark{6},     X.Y.~Zhang\altaffilmark{6}, 
        Zhaxiciren\altaffilmark{3}, and 
        Zhaxisangzhu\altaffilmark{3}\\
         (The Tibet AS${\bf \gamma}$ Collaboration)}

\altaffiltext{1}{ Department of Physics, Hirosaki University, Hirosaki 036-8561, Japan}
\altaffiltext{2}{ Department of Physics, Saitama University, Urawa 338-8570, Japan}
\altaffiltext{3}{  Department of Mathematics and Physics, Tibet University, Lhasa 850000, China}
\altaffiltext{4}{ Laboratory of Cosmic Ray and High Energy Astrophysics, Institute of High Energy 
          Physics, Academia Sinica, Beijing 100039, China}
\altaffiltext{5}{  Department of Physics, South West Jiaotong University, Chengdu 610031, China}
\altaffiltext{6}{ Department of Physics, Shangdong University, Jinan 250100, China}
\altaffiltext{7}{  Faculty of Engineering, Kanagawa University, Yokohama 221-8686, Japan}
\altaffiltext{8}{  Faculty of Education, Utsunomiya University, Utsunomiya 321-8505, Japan}
\altaffiltext{9}{  Institute for Cosmic Ray Research, University of Tokyo, 
Kashiwa 277-8582, Japan}
\altaffiltext{10}{ Department of Physics, Konan University, Kobe 658-8501, Japan}
\altaffiltext{11}{ Faculty of Systems Engineering, Shibaura Institute of Technology, Omiya 330-8570, Japan}
\altaffiltext{12}{ Faculty of Engineering, Yokohama National University, Yokohama 240-0067, Japan}
\altaffiltext{13}{ Department of Physics, Yunnan University, Kunming 650091, China}
\altaffiltext{14}{ National Institue of Informations, Tokyo 101-8430, Japan}
\altaffiltext{15}{ Tokyo Metropolitan College of Aeronautical Engineering, Tokyo 116-0003, Japan}
\altaffiltext{16}{ Shonan Institute of Technology, Fujisawa 251-8511, Japan}
\altaffiltext{17}{ Solar-Terrestrial Environment Laboratory, Nagoya University,Nagoya 464-8601, Japan}

\begin{abstract}
We have shown that the Sun's shadow by high energy cosmic rays moves year by year and its behavior 
is correlated with  a time variation of the large-scale structure of the solar and interplanetary 
magnetic fields.
The solar activity was near minimum in the period from 1996 through 1997.
Using the data obtained with the Tibet air shower array, we examined the shadowing of cosmic rays by the Sun in this quiet phase of solar cycle, and found that
the Sun's shadow was just in the apparent direction of the Sun, though it was observed at the 
position considerably away from the Sun
to the south-west in the period between 1990 and 1993. It is known that the magnetic pole 
of equivalent solar dipole was reversed during the previous active phase, and near solar
minimum the dipole was aligned with the rotating axis,  preserving  its N-pole  on the north pole side of the Sun.
This causes the solar magnetic field to shift the Sun's shadow to the east. Thus,
the observed results suggest that the shift of the Sun's shadow due to the solar magnetic field
was pushed back by the effect of the geomagnetic field, since the geomagnetic field always make the shadow
shift to the west. We discuss the Sun's shadow observed
during the period near solar minimum in 1996-1997 and compare it 
with the simulation results.

\end{abstract}


\section{INTRODUCTION}

  The Sun casts the shadow in the galactic cosmic-ray flux 
coming from the direction of the 
Sun. As almost all cosmic rays are positively charged particles, they are somewhat bent by the
magnetic fields between the Sun and the Earth if their energies are not so high 
and eventually this effect will make the Sun's shadow shift to
a position away from the apparent Sun's direction.
This effect was  first observed with  the Tibet-I air shower array, operating at
 Yangbajing in Tibet (4,300m above sea level) since 1990, in 1993 (\cite{ame93a}). We have also
shown that the displacement of the
Sun's shadow is strongly correlated with the cycle of solar activity (\cite{ame93b} ; \cite{ame96}).

  It is  known that the 
interplanetary magnetic
field (IMF) is formed as a result of the transport of the photospheric magnetic field
by the solar wind flowing continuously from the Sun (\cite{park63}).
Field lines near the solar equator form closed loops (neutral sheet), 
while field lines
from the poles are dragged far into interplanetary space by the high
speed solar wind of about 800 km/s. Furthermore, the magnetic field of the Sun
changes sign from south  to north  across the neutral sheet.
  Accordingly the IMF is organized 
into large regions of opposite polarity separated by the neutral sheet, which is wavy and is inclined to
the plane of the ecliptic. Then, the large-scale structure 
of the warped neutral sheet causes a sector structure of the IMF 
with the field direction reversing 
across the sector boundary (\cite{wilcox65})  so that
the magnetic filed points inward in some sectors and outward in others. 
This sector structure (``toward'' and ``away'') observed in the IMF, however,  
would vary with the phase of the solar activity cycle. The existence of 
the neutral sheet has been well-established observationally 
 since the work by Smith  et al. (\cite{smith78}).

  The IMF structure  can  be explained fairly well 
by the so-called  ``ballerina skirt'' model 
assuming a rotating dipole in the 
Sun (\cite{schult73} ; \cite{saito75} ; \cite{svalg78}), though direct evidence for
the presence of such rotating dipole  has 
not been obtained yet. In this model the variations and the reversal 
of the polar fields 
are  characterized
in terms of the rotating dipole. 
Near solar minimum the dipole is aligned with the rotating axis. 
Rotating slowly, the dipole becomes equatorial near maximum
and finally approaches the rotation axis again near the following minimum, 
but pointing in the opposite direction. This model also predicts
a planar neutral sheet with relatively small warps near solar minimum.

The {\em Ulysses} mission  explored the magnetic field of the Sun's 
south polar region at the
radial distance of about 2.3 AU from the Sun being at near solar minimum
activity  (\cite{smith95} ; \cite{fisk97}).  
The radial component of
the magnetic field is most easily related to the global solar magnetic field.
If the solar magnetic field is of dipolar form, its magnitude is expected to vary
with latitude and the field strength should show a significant
increase over the polar region. Actually many models of the solar 
magnetic field
discussed in the past had assumed a field similar to
that of a dipole.  {\em Ulysses}, however, found that the radial 
component of the magnetic field
was independent of latitude (\cite{smibalo95} ; \cite{balo95}), 
that is, the dipolar pattern of the field strength
is not reflected in the heliospheric magnetic field.   It is known that the magnetic 
field in the high speed solar wind originates in the polar
coronal hole occupying a limited region near the pole of the Sun.
 Then, the {\em Ulysses} data
may suggest that a
 dipolar pattern close to the surface of the Sun becomes a uniform pattern
at the location where the solar wind begins to flow radially outward.

Fisk  discussed a model to interpret the {\em Ulysses} data  (\cite{fisk96}).
This model assumes that the magnetic field in the 
inner corona of the polar corona 
hole can be represented by a tilted dipole,
which has a magnetic axis offset from the solar rotation 
axis, and which rotates rigidly at the equatorial rotation speed.
The field lines are anchored in the photosphere,  which differentially rotates,
while corona holes rotate rigidly at near the equatorial rotation rate.
Then a field line will move through the corona hole pattern and experience a variety
of non-radial expansion in the corona hole. 
As the field lines open into the heliosphere, they expand in heliomagnetic latitude so that
the magnetic field strength that is dipolar near the solar surface becomes
uniform in the heliosphere.  

In any case, as discussed above, the dipolar field of the Sun plays a major role
 for the origin of the IMF structure. In the previous papers
(\cite{ame93b} ; \cite{ame96}), we showed that the Sun's shadow is sensitive to
the magnetic field close to the Sun ($< 10$ times the solar radius), where any spacecraft mission
 can not approach. Hence the continuous observation of the
Sun's shadow over the cycle of solar activity may reveal a great deal 
about the behavior of a 
rotating dipole in the Sun, if it is in existence. 
 
When the Sun is in a quiet phase, the solar magnetic field is 
symmetric between 
the northern  and southern  hemisphere of the Sun.  Certainly {\em Ulysses}
 in 1994 observed
nearly the same magnetic field in both hemispheres.
With increasing solar activity, however,
the number and the activity of solar active regions tend to increase and these active
regions disturb the configuration of the solar magnetic field. This effect should be reflected
on the Sun's shadow. 
Actually, the Sun's
shadow was observed in the direction shifted considerably from the Sun's 
direction to the south-westward in 1990-1993 (\cite{ame93a} ; \cite{ame96}).
This period was just in the active phase of solar cycle, so the observed displacement
can be attributed to  a change of the solar magnetic field. 
The solar activity, however,  was near minimum in 1996-1997.  During this period, then, 
we may be able to observe a direct effect of the equivalent solar 
dipole field on 
the Sun's  shadow. We examine this for the first time in this paper.

\section{EXPERIMENT}

The Tibet-I air shower array (\cite{ame92})  was updated by increasing the number of counters in 1995. Since then this new air shower array (Tibet-II) has been successfully operated. 
The Tibet-II array, consisting of 221 scintillation counters each being placed on a lattice of 15 m spacing
with a covering area of 36,900 m$^2$, has been triggering the air shower events at a rate of about 200 Hz under any 4-fold coincidence in the detectors. Its effective area is about 7 times as large as the Tibet-I array.

 We analyzed the data set taken during the period 
from 1995 October through 1997 August.
 The event selection was made as in our  previous
analysis (\cite{ame93a}).
For the analysis of the shadows of the Moon and the Sun, we further selected the events 
within a circle of the radius $8^\circ$ of the Moon and the Sun with the zenith angle
less than 45$^\circ$.
 A coordinate system was 
fixed on the object, putting the origin of
coordinates on its center.  The position of each event observed is
then specified by the angular distance $\theta$ and the position angle
$\phi$, where $\theta$ and $\phi$ are measured from the center and
from the north direction, respectively. In the following, we chose 
the equatorial coordinates for the Moon and the ecliptic coordinates for
the Sun, respectively. The event distributions plotted in such
coordinate systems are used to examine the shadowing of cosmic rays by
these objects (\cite{ame93a}).

In this paper, our analysis is focused on the Sun's shadow observed with the Tibet II array during the period 
from 1995 through 1997, which almost corresponds to near minimum  of the 
cycle of solar activity changing with a 11 year period. The Ulysses 
result (\cite{fisk97}) shows that during the current cycle the polarity is outward 
in the northern  hemisphere and inward in the 
southern hemisphere, that is, N-pole of the solar dipole is in the northern hemisphere.

\section{MOON'S  SHADOW AND GEOMAGNETIC FIELD}

As discussed in the previous paper (\cite{ame93a}), the Moon's shadow provides  a good estimate of the angular resolution and systematic pointing accuracy of the air shower array at high energies. 
The angular resolution of this  array has been confirmed to be better than 1$^\circ$ by observing the Moon's shadow (\cite{ame93a}), and the mode energy of primary cosmic rays responsible for generating air showers detected by this array is estimated to be about 8 TeV for protons.

It is also noted that the Moon's shadow and the geomagnetic field, 
which acts  as a convenient momentum analyzer,
enables us to calibrate  the results obtained by assigning primary energies to all 
air shower events. Actually, the deflection angle of a proton of energy $E$ by the 
geomagnetic field  is  calculated as 
	$\Delta\theta = 1.6^\circ \times (E /1 {\rm TeV})^{-1}$ and 
the direction of this field should make the shadow shift to the west.

We first examine  a  correlation between the deflection angle of the Moon's shadow and the
air shower size  using the events observed with the Tibet-II array.  For this,  we searched 
for the  deficit center of the Moon's shadow by dividing the events into  four size regions  of 
$15 < \sum\rho < 50$, $50 < \sum\rho < 100$, 
$100 < \sum\rho < 300$ and  $300 < \sum\rho$, 
where $\sum\rho $ is the sum of the number of particles observed in each detector. 
Using the Monte Carlo simulation, the mean  energies  of the primary particles generating 
the events fallen 
 in respective size regions are estimated to  be about 8~TeV, 15~TeV, 35~TeV and  100~TeV, respectively.

\begin{figure}
\epsscale{.7}
\plotone{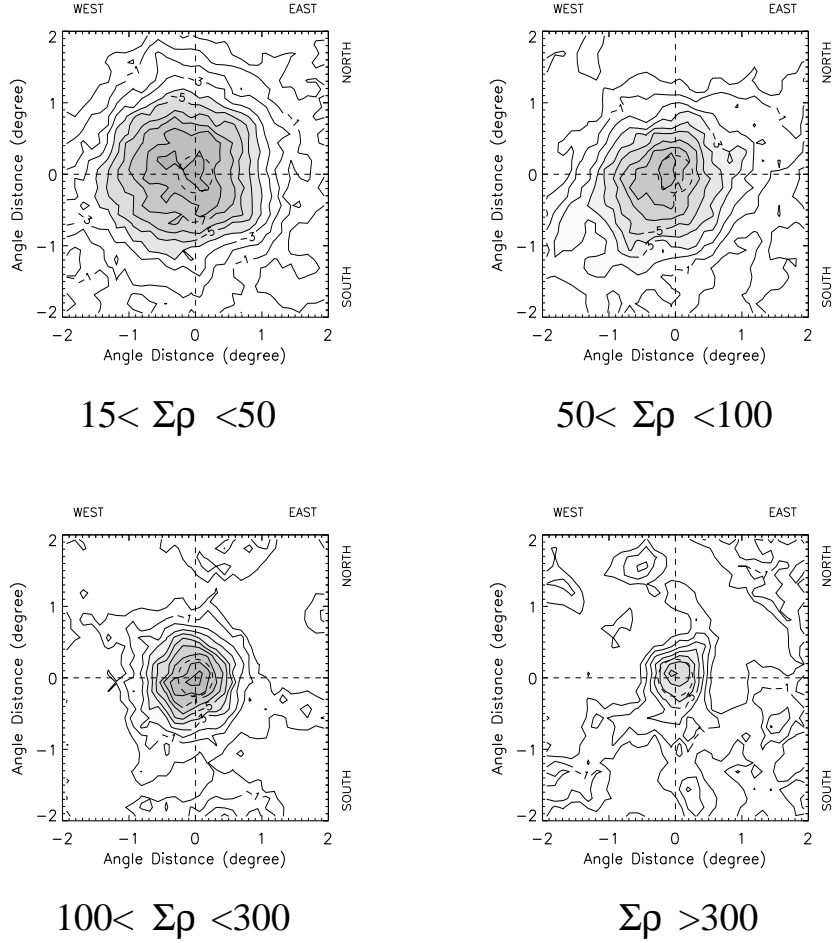}
\caption {Energy dependence of the displacement of the Moon's shadow.
          The contour map of each shadow gives
 	the weight of deficit event density from the background, and contour
	lines start from 0$\sigma$ deficit with a step of 1$\sigma$. Same definition
        is used in the following figures.  \label{Fig1}}
\end{figure}

Figure \ref{Fig1} shows the event density map of the Moon's shadow in each size
 region, obtained using  the events with $\sum\rho >15$
 observed during the period from 1995 to 1997.  
It is seen  that the center of the Moon's shadow is observed in the direction shifted to the west in
small shower size region.

The positions of the most deficit center in respective size intervals  are estimated to be 
${0.21^{\circ}}^{+0.15^{\circ}}_{-0.18^{\circ}}$W, 
${0.18^{\circ}}^{+0.19^{\circ}}_{-0.13^{\circ}}$W, 
${0.04^{\circ}}^{+0.07^{\circ}}_{-0.08^{\circ}}$W and 
${0.04^{\circ}}^{+0.08^{\circ}}_{-0.06^{\circ}}$E, respectively.  
Here, a maximum likelihood method was used by assuming a 2-dimensional Gaussian type 
probability function for the event density. 

\begin{figure}
\epsscale{.7}
\plotone{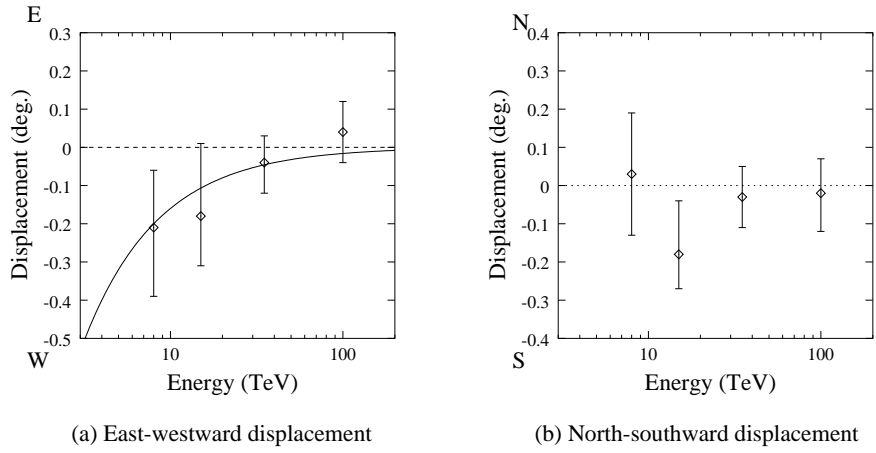}
\caption {Energy dependences of the displacements of the Moon's shadow in the east-west
        direction (a) and  north-south direction (b), respectively.
        Solid curve in (a) shows the expected deflection angle of a proton by 
	the geomagnetic field.    \label{Fig2}}
\end{figure}

Figure \ref{Fig2} (a) and (b) show the plots between 
the deflection angle and the mean primary energy in the east-west direction and
north-south direction, respectively. As shown in Figure \ref{Fig2} (a), the 
east-westward displacement of the Moon's shadow
is consistent with that expected from the effect of geomagnetic field, while 
Figure \ref{Fig2} (b) provides an estimate  of the  pointing accuracy of 
the air shower array, which is smaller than 0.1$^\circ$.

\section{YEARLY VARIATION OF THE SUN'S SHADOW}

Table \ref{IMF_power} shows a yearly variation of the magnitude of solar and interplanetary magnetic fields taken from the data of Stanford Mean Solar Magnetic Field (\cite{noaa}) and IMP8 (\cite{nasa}). 
This table tells us that the strength of IMF near the Earth is rather stable despite 
changes of solar activity,
but the mean strength of solar magnetic field reached a maximum around 1991 and then decreased rapidly with decreasing of the solar activity. This data also shows that the cycle 
of solar activity reached near minimum between 1996 and 1997.

\begin{figure}
\epsscale{.7}
\plotone{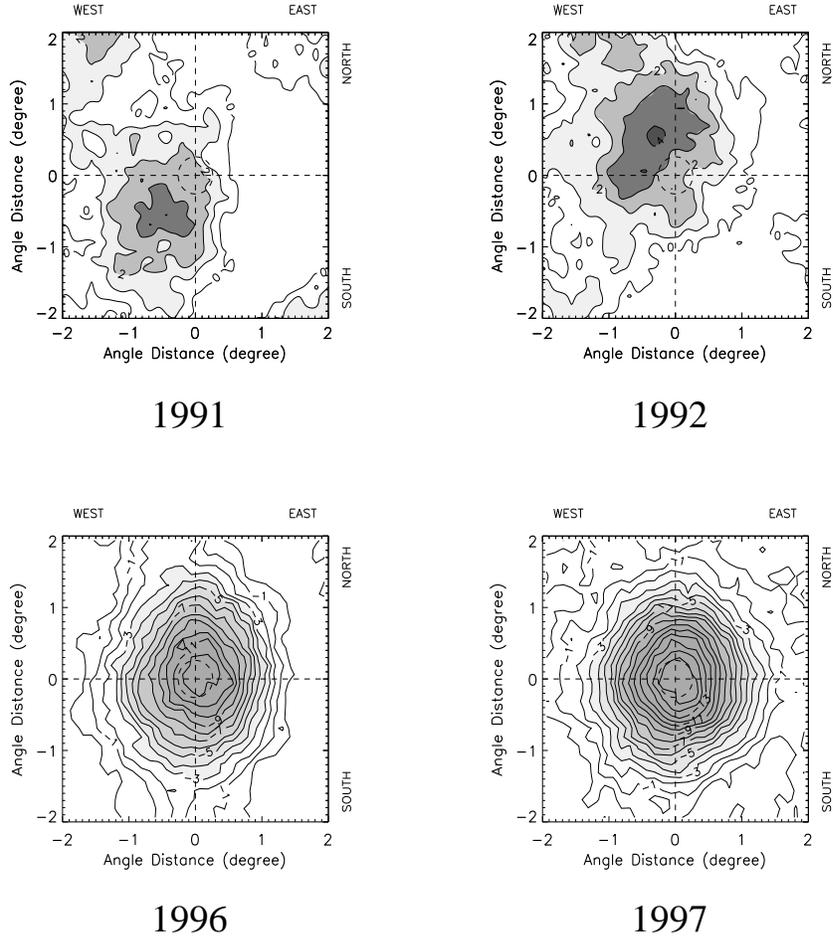}
\caption { Yearly  variation of the Sun's shadow. Each shadow is obtained using
 all events with $\sum\rho > 15 $ and the mode energy of primary
protons is estimated to be about 8 TeV.   \label{Fig3}}
\end{figure}

Shown in  Figure \ref{Fig3} is a  yearly variation of the Sun's shadow 
 observed with the Tibet-I and Tibet-II arrays. The observation  period for each shadow is as follows :  1991 (1991.4-1991.8) ; 
1992 (1992.3 - 1992.7) ; 1996 (1996.3-1996.8) and 1997 (1997.3-1997.8).  
As discussed in detail in the previous paper (\cite{ame96}), a considerable change is
  observed between 1990 and 1993 corresponding to the period near  maximum or at
declining phase of solar activity, and its movement seems to become quick steps after 
that. It is observed  that during the active phase  the  Sun's magnetic dipole
changed its direction often and abruptly  and finally its field polarity was completely
 reversed. 
A sudden change of the Sun's polar fields  that caused an asymmetry of the away and toward sector field would create such a large displacement of the shadow in 1991-1993.
Therefore, the Sun's shadow correlates strongly with a variation of large scale structure of the solar magnetic field changing  with the phase of solar cycle.

Compared with the data in 1991 and 1992, however, the Sun's shadows 
observed in 1996 and 1997 were found almost at the apparent solar 
position. In the following we discuss the behavior of the Sun's shadow observed in the quiet phase of solar activity.

\section{SUN'S SHADOW NEAR SOLAR MINIMUM }

\begin{figure}
\epsscale{.7}
\plotone{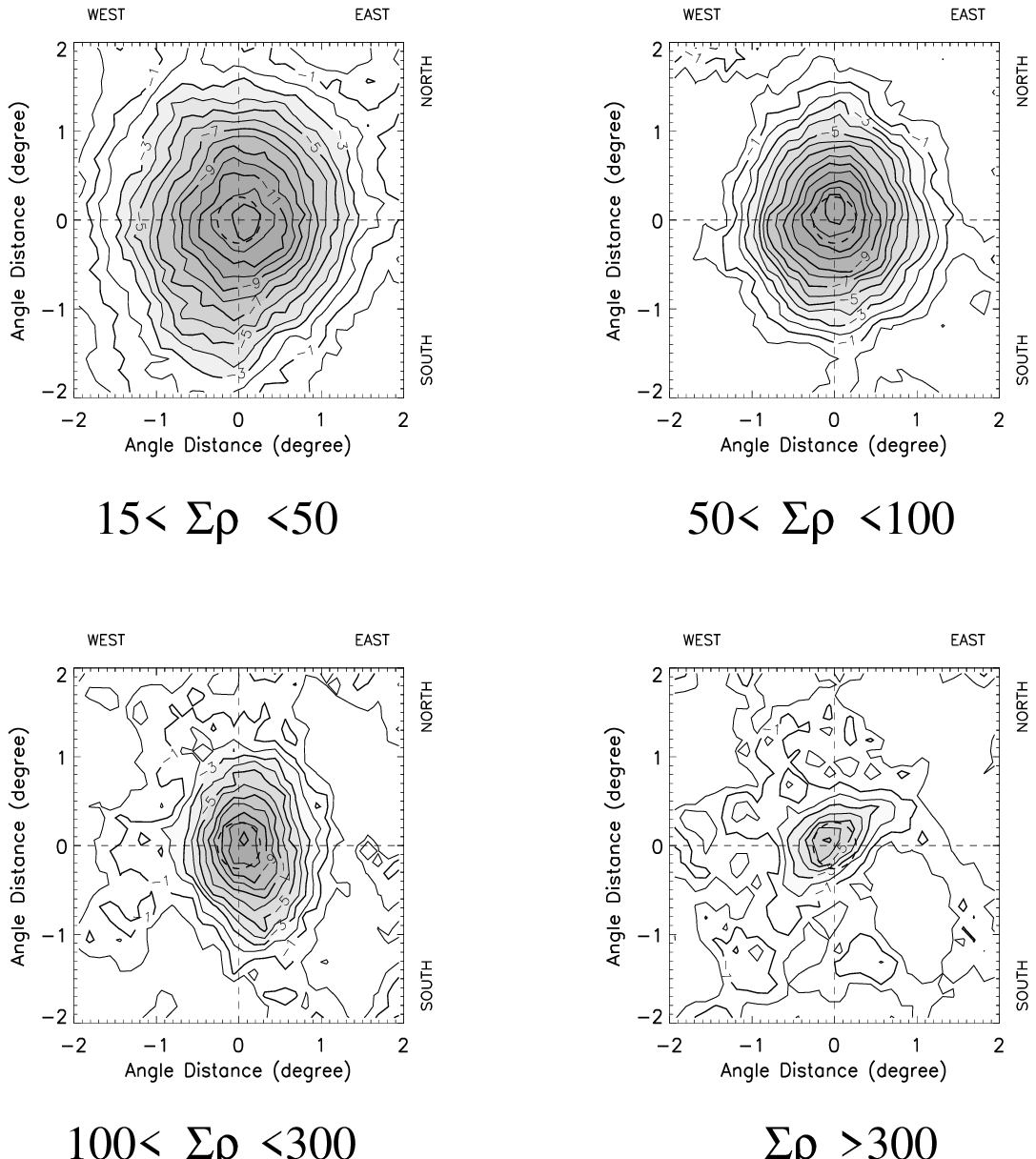}
\caption { Dependence of the Sun's shadow upon the air shower size, observed in 1996.  \label{Fig4}}
\end{figure}

In Figure \ref{Fig4}, we show the event density maps of the Sun's shadow 
in four different size regions, which were observed with the Tibet-II array in 1996.
Each size interval is  the same as that in Figure \ref{Fig1}.
The observed displacement of each Sun's shadow in this figure is a 
superposition of
the effects of the solar magnetic field and the geomagnetic field. 
It is of a great interest to note that the shadow of each size region was observed almost in the 
apparent direction of the Sun, independent upon the air shower size or primary energy.  This suggests
that the effects of the solar magnetic field and the geomagnetic field may be canceled each other,
resulting in that the shadow remained just in the Sun's direction.
In order to see this in detail, we examined separately an energy dependence of the
displacement of the shadow in the away and toward sectors.  
For this, the Tibet-II data set was divided into three subsets  according to 
the away field (field strength $>$ 0.5 nT), toward field ($<$ -0.5 nT) and boundary
 field (between -0.5 nT and 0.5 nT) using the IMP8 data (\cite{nasa}). 
The fraction of each data set is 3.8 : 3.3 : 2.9. 

\begin{figure}
\epsscale{.7}
\plotone{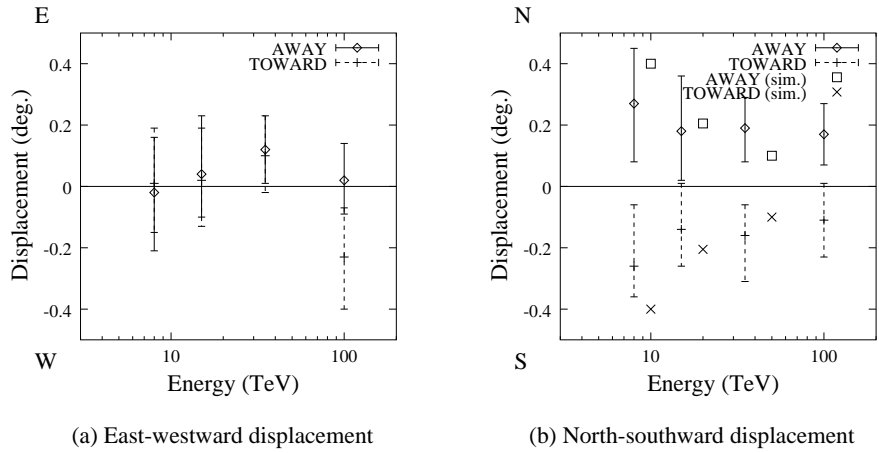}
\caption {Energy dependence of the  east-westward (a) and  north-southward (b) 
         displacement  of the Sun's shadow.  Minus and cross markers denote the  displacement 
	of the Sun's shadow for the away IMF and toward IMF data sets, 
	respectively. The simulation results 
        (squares : away ; crosses : toward) 
          are compared with the experimental data. \label{Fig5}}
\end{figure}

 The displacement of the shadow in the east-west direction
is shown in Figure  \ref{Fig5}(a) for both the away and toward sectors. No obvious 
displacement  is seen for the shadow in each sector, independent upon the primary energy.
For the north-southward displacement, however, the Sun's shadow in each sector of the IMF
is  shifted in the opposite direction according to the direction of
the sector field, and its displacement is symmetric to the ecliptic plane,
as seen in  Figure \ref{Fig5}(b).  The Sun's shadow was not clearly observed in the
boundary field  because of non-uniformity of the field direction. 

  It is known that near solar minimum the solar magnetic field is symmetric between 
the north and 
south hemisphere of the Sun and the neutral sheet is laid on the ecliptic plane
with relatively small warp.
The IMF near the Earth is almost parallel to the ecliptic plane and has  an Archimedian spiral configuration (\cite{park63}). 
The azimuthal component of this  Parker field becomes dominant at large distance from the Sun. Typical crossing angle of this field and Earth's orbit is about 45$^\circ$. The field strength decreases inverse proportion to the distance. Average value of this field observed by IMP8 during the period from 
1996 to 1997 is  3.4 nT near the orbit of the Earth. The field component at right angles
to the cosmic ray trajectories will make the shadow shift to the north in the away sector and to the south in the toward sector. 
Furthermore, the geomagnetic field shifts the Sun's shadow
to the west only, since the direction of the Earth's magnetic dipole is almost parallel to the  rotation axis of the Earth.
 Therefore, the  north-southward displacement of the shadow observed in the away and toward
sectors can be mostly attributed to the effect of the IMF.

On the other hand, the IMF will scarcely make the Sun's shadow shift to the east-westward
 because of a symmetric  distribution of the away and toward field to the ecliptic 
plane in the period near solar minimum. Therefore, the displacement in the east-west
direction is, for the most part, caused by the effects of the solar dipole field and the geomagnetic
field. The observed results  shown in Figure  \ref{Fig5}(a) suggest that
the eastward displacement of the shadow  by the solar dipole field was
seemingly pushed back by the westward displacement due to the geomagnetic field so that
no displacement in the east-west direction was found for the Sun's shadows observed 
in the away and toward sectors.

The behavior of the shadow observed in 1997  is almost same as in 1996. 
We expect that the Sun's shadow in each  sector should show a different 
behavior according to the
change of solar activity as in 1991-1993, and this may  be further checked 
with better statistics during the present solar cycle 23. 

\section{COMPARISON WITH A SIMULATION}

A calculation of the shadows of the Sun and the Moon was made 
 using a simple model of 
the solar and interplanetary magnetic field  (\cite{suga99}).
 The assumptions made on the magnetic fields are as follows,
\begin{enumerate} 
\item  For the geomagnetic field, a  dipole field  with the magnetic moment of $M_{E} = 8.1\times 10^{15}$ T$\cdot$m$^{3}$ is assumed. 
The axis of the dipole is inclined at 11.7 degrees from the rotation
axis of the Earth. 

\item For the solar magnetic field, a dipole field with the magnetic moment of $M_{S} 
=  B_0 R_\odot^3  = 1.7 \times 10^{22}$ T$\cdot$m$^{3}$ is assumed within a radius of about 10 $R_\odot$, where $B_0$ is the field strength at the surface of the Sun and 
$ R_\odot$  the solar radius. 
The dipole axis is parallel or anti-parallel to the rotation axis of the Sun,
that is, perpendicular
to the plane of the ecliptic.
The field strength of the dipole field at the radial distance $r$
from the Sun's center is expressed as
$ B = B_0 (R_\odot/r)^3 \sqrt{ 1 + 3 \sin^2 \Phi} $,
 where $\Phi$ is the angle measured from the plane of ecliptic.
 The strength of the dipole field decreases in proportion to
$r^{-3}$ and the magnetic moment gives the field strength of 100 $\mu$T at the polar surface ($\Phi=90^\circ$) and 50 $\mu$T at
 the equatorial surface of the Sun ($\Phi=0^\circ$), respectively.
\item  The IMF is continuously  generated  by the steady solar wind 
blowing from the
corona with a constant velocity of $v_r$ = 450 km/s. The solar wind spreads 
two-dimensionally in the 
plane of the  ecliptic and the IMF has an Archimedean spiral configuration as the Sun rotates.
 The radial and azimuthal components of the IMF at the distance $r$
from the Sun are expressed as $ B_r = B_0 (R_\odot/r)^2$ and
 $B_\phi \simeq - B_0 (R_\odot /r)^2(r \omega_0/v_r) (r>>R_\odot)$
near the equatorial plane, respectively, where  $\omega_0$ is the rotation angular velocity of the Sun. 
Then, 
$ B = \sqrt{B_r^2 + B_\phi^2} \simeq B_0 (R_\odot/r)^2 \sqrt{1 + r^2\omega_0^2/v_r^2} $. 
It is seen that the field strength decreases inversely proportional  
to $r$ when the distance $r$ is very far from the Sun. 
Here, the magnitude of $B_\phi$ is taken to be 2 nT 
at the Earth's orbit  ( $r$ = 1 AU) with the garden 
hose angle 
of $45^{\circ}$. The magnitudes of solar dipole field and IMF take the
same value of 
40 nT at $r \simeq 10.8 R_\odot \equiv 0.05$ AU, at which the magnetic field is 
switched from the solar dipole one to the IMF, though the direction of line of 
force makes a jump. But, this jump has no effect on the result.

\item The IMF near solar minimum consists of four sectors, two away and two toward sectors.
Actually the observation data of the IMF (\cite{noaa} ; \cite{nasa}) show
  a  structure close 
to four sectors in the period from 
1996 through 1997. These are
axial symmetric on the plane of the ecliptic and make the Archimedian spirals.
The IMF changes its strength periodically across the sectors in the azimuthal direction.
We further assumed that its variation can be described in terms of the sin of azimuth
angle for a fixed distance $r$. The field strength becomes zero on the 
spiral boundaries between the away and toward sectors and its absolute value reaches  the maximum at
the center of each sector for the fixed distance $r$.  The root mean square value of the field strength
averaging over the azimuth angle is taken to be equal to the value discussed above.

\end{enumerate}

Of course, the solar magnetic field near the Sun is
more complicated than the dipole as found by {\em Ulysses}.
 It is, however,  known that the polarity
of the field is opposite in the north and south across  the neutral sheet,  which is
almost parallel to the plane of ecliptic near solar minimum.
In this sense, the global field structure near the Sun is
  similar to that of a dipole, while the high-latitude magnetic
field may deviate from the expected dipole geometry according to the
{\em Ulysses} observation (\cite{smibalo95} ; \cite{balo95}).
 Also, the Sun's shadow near solar minimum is not so sensitive to the
 field components in the polar direction. Hence, our model 
for the solar magnetic field may be a good enough approximation for the present work.
Furthermore,  such  
simplification is very helpful in  understanding  how the Sun's shadow 
is affected by the dipole component of the solar magnetic field.

Trajectories of charged particles are calculated by emitting anti-charged particles 
from Yangbajing in the wide angular range of 25$^\circ \times 25^\circ$  in real angle 
around the Moon and the Sun. The simulation results were re-sampled by taking into account the energy spectra of
primary particles and the angular resolution of the Tibet-II array. 
Primary particles are assumed to be composed of protons and helium nuclei with a power-law spectrum of $E^{-\gamma}$, where
$\gamma$ = 2.7 for both particles, since other nuclei give a minor 
contribution.
 The flux ratio of protons to helium nuclei is taken to be 1.5. Trigger efficiencies of the events induced by primary cosmic rays were calculated by the simulation.

\begin{figure}
\epsscale{.7}
\plotone{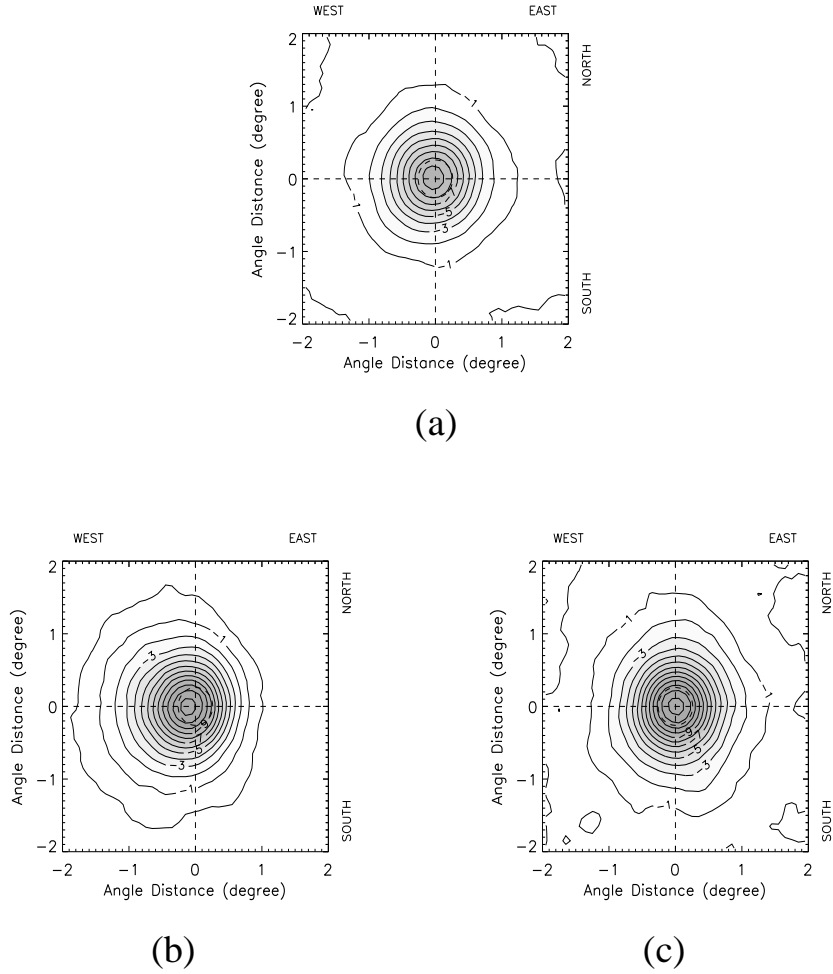}
\caption {Shadows of the Moon and the Sun  obtained by the simulation 
   (all events with $\sum\rho >15$).   (a)  Moon's shadow, (b)  Sun's shadow in case of 
           parallel dipoles and  (c) Sun's shadow in  case of 
           anti-parallel dipoles.  \label{Fig6}}
\end{figure}

Figure \ref{Fig6} shows the shadows of  the Moon and the Sun  obtained
 by the simulation under the same observation conditions as the experiment. 
The Moon's shadow shown in Figure \ref{Fig6}(a) is consistent 
with the experiment, which is shifted to the west by about 0.15$^\circ$. For the Sun's shadow,
we examined two cases that the dipole directions of the Sun and the Earth are
 (1)  parallel and (2) anti-parallel, respectively. In  case of (1) the displacement
by the Sun's dipole is the same direction as that by the Earth's dipole, while
in  case of (2) the direction of each displacement becomes opposite. 
The simulation results by both cases are
shown in Figure \ref{Fig6}(b) and (c),
respectively.  The case (2) may correspond to the shadow observed during 
the period from 1996 to 1997, and it is consistent with the experiment.
This simulation may suggest that the most suitable value, $M_s$, for the magnitude of
the Sun's magnetic moment can be estimated from a  detailed analysis of
 the Sun's shadow near solar minimum.
   
We also examined the displacement of the Sun's shadow by the IMF. The simulation results (\cite{suga99}) show
that the shadow is shifted to the north by the away IMF and to the south
by the toward IMF, which are consistent with the experiment as shown in Figure \ref{Fig5}(b).
Because of a difficulty of separating the away and toward field from the observation data,
the observed displacements of the shadow in each sector are slightly smaller than the
simulation results.  
It may be also seen that the Sun's shadow shown in Figure \ref{Fig4} 
shows an oval in shape with the major axis in the 
north-south direction. This deformation can be explained by the effect of
IMF.

A comparison with the simulation and the experiment may strongly support the following :
1) Near solar minimum, the displacement of the Sun's shadow in the east-west direction
is mostly caused by the effect of the dipole component of the solar 
magnetic field ;  and 2) the displacement in the north-south
direction is due to the effect of the IMF. In the last solar minimum, the effect
of the dipole field of the Sun was almost canceled by the geomagnetic
field so that the Sun's shadow seemingly stayed almost in the Sun's direction.  In the next solar minimum,
however, the Sun's shadow will be shifted to the west as  in
Figure \ref{Fig6} (b) since the polarity of the
Sun's dipole is reversed again by that time (case(1)).

\section{SUMMARY}

 Using the data obtained with the Tibet air shower array during the period from 1995 October to 1997 August, 
we examined the behavior of the Sun' shadow and compared it with a simulation. In this period, the Sun's shadow
was observed almost in the apparent Sun's direction, and shows a remarkable difference compared  with
those observed in 1990 - 1993.   It is confirmed that the polarity of the Sun's dipole was  reversed in 1992 - 1993
and since then  the Sun's dipole has been in the opposite direction to the Earth's dipole (anti-parallel).
From a comparison of the observed shadows and the simulation results using a model of the solar and interplanetary
magnetic field, we confirmed  that the eastward displacement of the shadow due to the solar magnetic field
 was almost forced back by the effect of the geomagnetic field. We also examined that during this
period the IMF made the shadow shift to the north in the toward sectors and to the south in the away sectors.  
A detailed analysis of the Sun's shadow near solar minimum may enable us to 
estimate a magnitude of the magnetic moment of the Sun's (equivalent) dipole. 

The Sun is now going toward the active phase of the cycle 23 
and will reach a maximum around the year of 2001.
Then, the angle of the dipole component should  rotate from north to south during the cycle, 
reversing near maximum. The behavior of a dipole may be directly examined from the observation
of the Sun's shadow during the period near maximum or at declining phase of 
solar cycle with a two sector structure of the IMF polarity.
During the coming active phase,  the Sun's shadow would
 be found in the direction fairly away from the Sun's direction  as in 1990 - 1993,
but possibly shifted to the north-eastward,
  though its direction and magnitude may depend upon the dominance of the dipole component.

A new Tibet air-shower array will be operated in very near future.
  This array consists of 545 scintillation
detectors, which are placed on a 7.5 m grid with a covering area of 2 $\times 10^4$ m$^2$. 
Using this array, air shower events with energy in excess of  about 3 TeV can be detected with no serious bias.
Then, we can study  the Sun's shadow in a wide energy range from 3 TeV to 100 TeV
with better statistics. A further observation of the Sun' shadow  may provide
a unique and important means for clarifying a dynamical change or structure
of the solar and interplanetary magnetic fields close to the Sun  
and also provide new clues how to study the three dimensional configuration of the
solar magnetic field under the influence of the solar activity changing 
with a 11 year period.

\acknowledgments

  This work is supported in part by Grants-in-Aid for Scientific
Research and also for International Science Research from the Ministry
of Education, Science, Sports and Culture in Japan and for International Science Research from
the Committee of the Natural Science Foundation and the Academy of Sciences in China.
The authors thank the anonymous referee for critical and valuable comments.

\clearpage

\begin{table*}[t]
\begin{center}
\begin{tabular}{|c|c|c|c|c|c|c|c|c|c|}
\hline
Year   & 1990  & 1991  & 1992  & 1993  & 1994  & 1995  & 1996 & 1997 & 1998\\ \hline
Stanford ($\mu$T)    & 47.0  & 72.1  & 34.1  & 26.4  & 23.6  & 14.7  & 9.1 & 8.9 & 17.6\\ \hline
IMP8 (nT)      & 5.1   & 6.4   & 5.7   & 4.4   & 4.3   & 3.7   & 3.1 & 3.7 & 3.7\\ \hline
\end{tabular}
\end{center}
\caption{
 Yearly variation of the mean strength of solar magnetic field and IMF near the Earth.  \label{IMF_power}}
\end{table*}

\clearpage

\clearpage

\clearpage

\clearpage

\clearpage

\clearpage


\begin{thebibliography}{}
\bibitem[Amenomori et al.\ 1992]{ame92} Amenomori, M. et al. 1992, \prl, 69, 2468
\bibitem[Amenomori et al.\ 1993a]{ame93a} Amenomori, M. et al. 1993a, \prd, 47, 2675
\bibitem[Amenomori et al.\ 1993b]{ame93b} Amenomori, M. et al. 1993b, \apj, 415, L147
\bibitem[Amenomori et al.\ 1996]{ame96} Amenomori, M. et al. 1996, \apj, 464, 954
\bibitem[Balogh et al.\ 1995]{balo95} Balogh, A. et al. 1995, Science, 268,
1007.
\bibitem[Fisk\ 1996]{fisk96} Fisk, L.A. 1996, J. Geophys. Res., 101, 15, 547.
\bibitem[Fisk\ 1997]{fisk97} Fisk, L.A. 1997, Proc. 25th Int. Cosmic-Ray Conf. (Durban),
 8, 27
\bibitem[NASA/NSSDC]{nasa}  NASA/NSSDC, National Space Science Data Center(NSSDC), {\em OMNIWeb - Near Earth Heliosphere Data} (URL http://nssdc.gsfc.nasa.gov/omniweb/)
\bibitem[NOAA/NGDC]{noaa}  NOAA/NGDC, National Geophysical Data Center(NGDC), {\em Solar and Upper Atmospheric Data Sources} (ftp://ftp.ngdc.noaa.gov/STP/SOLAR\_DATA/SUN\_AS\_A\_STAR/STANFORD/)
\bibitem[Parker\ 1963]{park63} Parker, E.J. 1963, Interplanetary Dynamical Process (New York : Interscience)
\bibitem[Saito\ 1975]{saito75} Saito, T.  1975, Sci. Rep. Tohoku Univ., Ser. 5, 23, 37
\bibitem[Schultz\ 1973]{schult73} Schultz, M. 1973, Astrophys. Space Sci. 24, 371
\bibitem[Smith et al.\ 1978]{smith78} Smith, E.J. et al. 1978, J. Geophys. Res. 83, 717.
\bibitem[Simth \& Balogh\ 1995]{smibalo95} Smith, E.J. \& Balogh, A. 1995,
J. Geophys. Res. Lett., 22, 3317.
\bibitem[Smith et al.\ 1995]{smith95} Smith, E.J., Marsden, R.G. and Page, D.E. 1995, Science, 268, 1005.
\bibitem[Suga et al.\ 1999]{suga99} Suga, Y. et al. 1999, Proc. 26th Int. Cosmic-Ray Conf. (Salt Lake City), 7, 202
\bibitem[Svalgaad \& Wilcox\ 1978]{svalg78} Svalgaard, L. \& Wilcox, J.M.  1978, Ann. Rev. Astron. Astrophys.
16, 429, 1978
\bibitem[Wilcox \& Ness\ 1965]{wilcox65} Wilcox, J.M. \& Ness, N.F. 1965, J. Geophys. Res., 70, 5783
\end{thebibliography}
\end{document}